\documentclass[aps,prb,twocolumn,secnumarabic,amsmath,amssymb,superscriptaddress]{revtex4-1}

\usepackage{color}
\usepackage{graphicx}
\usepackage{amsmath}
\usepackage{xspace}
\usepackage{bm}
\usepackage{times}
\newcommand{\MnFeGe}{Mn$_{1-x}$Fe$_x$Ge}

\begin{document}

\title{
	Control of Dzyaloshinskii-Moriya interaction in \MnFeGe: a first-principles study
}

\author{Takashi Koretsune}
\affiliation{RIKEN Center for Emergent Matter Science (CEMS), Wako, Saitama 351-0198, Japan}
\author{Naoto Nagaosa}
\affiliation{RIKEN Center for Emergent Matter Science (CEMS), Wako, Saitama 351-0198, Japan}
\affiliation{Department of Applied Physics, University of Tokyo, Hongo, Tokyo 113-8656, Japan}
\author{Ryotaro Arita}
\affiliation{RIKEN Center for Emergent Matter Science (CEMS), Wako, Saitama 351-0198, Japan}

\date{\today}

\begin{abstract}
	Motivated by the recent experiment on the size and helicity control of skyrmions in \MnFeGe\ [K. Shibata et al., Nature Nanotechnology {\bf 8}, 732 (2013)], we study how the Dzyaloshinskii-Moriya (DM) interaction changes its size and sign in metallic helimagnets.
	By means of first-principles calculations, we successfully reproduce the non-trivial sign change of the DM interaction observed in the experiment. 
	While the DM interaction sensitively depends on the carrier density or the detail of the electronic structure such as the size of the exchange splitting, its behavior can be systematically understood in terms of the distribution of anticrossing points in the band structure.
	By following this guiding principle, we can even induce gigantic anisotropy in the DM interaction by applying a strain to the system.
	These results pave the new way for skyrmion crystal engineering in metallic helimagnets.
\end{abstract}

\maketitle

A skyrmion is a topologically protected nano-size spin texture found in several magnets\cite{bogdanov1989,rossler2006,muhlbauer2009,yu2010,yu2011}.
Due to its unusual spin structure, many intriguing behaviors such as topological Hall effects, current-driven motion, and multiferroic behavior have been observed\cite{nagaosa2013}.
Although there is a huge potential to design novel functional materials by exploiting these unique electromagnetic properties, skyrmion engineering or skyrmion-crystal engineering is yet to be established.
Here, the key issue is how to manipulate the size and helicity of skyrmions.
Regarding this problem, a recent experiment for the representative skyrmion system \MnFeGe\ has shown that we can tune the skyrmion size and helicity by changing the carrier density\cite{shibata2013,grigoriev2013}.

The Hamiltonian which determines the nature of skyrmions is
\begin{align}
	H = \int d\bm r \left[ \frac{J}{2} (\nabla \bm M)^2 + D \bm M \cdot (\nabla \times \bm M) \right],
	\label{eq:Hamiltonian}
\end{align}
where $\bm M$ is the magnetization per volume, $D$ is the Dzyaloshinskii-Moriya (DM) interaction coefficient and $J$ is the ferromagnetic exchange coupling, respectively.
For materials design of skyrmion crystals, we need to know the precise value of $D/J$.
However, non-empirical evaluation of these parameters in the classical continuum model has been a difficult challenge, since it requires an elaborate multi-scale approach spanning the quantum to classical regime.
For the insulating skyrmion system Cu$_2$OSeO$_3$\cite{seki2012}, there is a work in which the spin Hamiltonian (1) was derived from first principles\cite{janson2014}.
By comparing the total energy of various magnetic states, they determined the value of $D/J$, and succeeded in reproducing the experimentally measured skyrmion size.
However, the guiding principle to control the values of $D/J$ is yet to be obtained. 

On the other hand, for metallic systems, the situation is different.
There are several studies for the estimate of $D$\cite{heide2008,ferriani2008,heide2009,udvardi2003,ebert2009,katsnelson2010,dmitrienko2014,freimuth2013,freimuth2014,wakatsuki2014}.
Among them, recently, one of the present authors (NN) and his collaborators have shown that the DM interaction in the two-band model drastically changes when the band anticrossing point resides near the Fermi level\cite{wakatsuki2014}.
The story is analogous to that of anomalous Hall conductivity, in which the band anticrossings act as magnetic monopoles in momentum space\cite{nagaosa2010}.
In Ref.~\onlinecite{freimuth2014}, a Berry phase expression for the DM interaction has also been formulated.
%These studies suggest that there is a fascinating possibility of controlling the DM interaction in metallic systems by manipulating the electronic structure.
These studies stimulate us to explore a fascinating possibility of controlling the DM interaction in metallic systems by manipulating the electronic structure.
Indeed, the fact that not only the size but also the helicity of skyrmions in \MnFeGe\ changes as a function of $x$ indicates that we have a good chance to control the value of $D$.

In this paper, we show a quantitative analysis of the DM interaction in the metallic helimagnet, \MnFeGe, based on {\it ab initio} density-functional theory (DFT) calculation.
From the obtained band structure, we evaluate the off-diagonal spin susceptibility which is a direct measure of the DM interaction.
We find that the sign change of $D$ observed in the experiment for \MnFeGe\ is successfully reproduced.
The carrier-density dependence of $D$ can be systematically understood in terms of the distribution of band anti-crossing points in the electronic structure.
We demonstrate that the sign and the size of $D$ can be controlled as a function of the carrier density or the size of the exchange splitting.
There is also an interesting possibility to induce gigantic anisotropy in $D$ by applying a strain to the system.\\

%\section{Results}

\noindent\textbf{\large{Results}}\\
%\subsection{Evaluation of DM interaction}
\noindent\textbf{{DM interaction in the continuum model.}}
Let us first look at the second term in the Hamiltonian \eqref{eq:Hamiltonian}. 
This indicates that $q$-linear term in the spin susceptibility, $\chi^{\alpha\beta}$, should be proportional to the DM interaction coefficient, $D$.
Therefore, to estimate $D$ in the continuum limit from the DFT calculation, we compute the long-wave length limit of the spin susceptibility, that is,
\begin{align}
	%\tilde{D}_\gamma \equiv - \lim_{q \to 0} \frac{\partial \chi^{\alpha \beta}(\bm q,i\omega_n=0)}{i \partial q^\gamma}.
	\tilde{D}_\beta \equiv \lim_{q \to 0} \frac{\partial \chi^{\alpha \gamma}(\bm q,i\omega_n=0)}{i \partial q^\beta}.
	\label{eq:DMcoef}
\end{align}
Here, $(\alpha,\beta,\gamma)=(x,y,z), (y,z,x),$ or $(z,x,y)$ and $\tilde{D}_\beta$ corresponds to the coefficient for $M_\alpha \partial M_\gamma/\partial \beta$ in Eq.~\eqref{eq:Hamiltonian}.
Since we consider the skyrmions in the $x$-$y$ plane under the total magnetic moment along the $z$-axis, hereafter we focus on $\tilde{D}_x$ and $\tilde{D}_y$.
In Eq.~\eqref{eq:DMcoef}, we use the non-interacting spin susceptibility defined as
\begin{align}
	%&\chi_0^{\alpha \gamma}(\bm q, i\omega_l) = \frac{T}{V} \sum_{l,l',s_1,s_2,s_3,s_4}\sum_{\bm k, m} \sigma^\alpha_{s_4 s_1}\nonumber\\
	%&\times G^0_{ls_1 l' s_2} (\bm k + \bm q, i\omega_m + i \omega_l) \sigma^\gamma_{s_2 s_3} G^0_{l' s_3 l s_4}(\bm k, i\omega_m)
	&\chi_0^{\alpha \gamma}(\bm q, i\omega_l) = -\frac{T}{V} \sum_{l,l',s_1,s_2,s_3,s_4}\sum_{\bm k, m} \sigma^\alpha_{s_4 s_1}\nonumber\\
	&\times G^0_{ls_1 l' s_2} (\bm k, i\omega_m) \sigma^\gamma_{s_2 s_3} G^0_{l' s_3 l s_4}(\bm k + \bm q, i\omega_m + i \omega_l)
\end{align}
where, $\sigma$ is the Pauli matrix and $G^0$ is the non-interacting Green's function in the orbital basis.
Using this non-interacting spin susceptibility, we can write as
$\tilde{D}_\beta = (1/V) \sum_{\bm k} \tilde{D}_\beta(\bm k)$
with
\begin{align}
	\tilde{D}_\beta(\bm k) &= \lim_{\bm q \to 0} \frac{\partial}{i \partial q_\beta} \sum_{n,n'}
	\frac{f(\varepsilon_{n' \bm k+ \bm q}) - f(\varepsilon_{n \bm k})}{\varepsilon_{n' \bm k+ \bm q} - \varepsilon_{n \bm k}}\nonumber\\
	& \times \langle n\bm k| \sigma^\alpha | n' \bm k + \bm q \rangle
	\langle n'\bm k+ \bm q| \sigma^\gamma | n \bm k \rangle,
	\label{eq:chiq}
\end{align}
where $| n \bm k \rangle$ is the eigenvector of the Kohn-Sham Hamiltonian with the eigenvalue of $\varepsilon_{n \bm k}$.
Hence, we can discuss the DM interaction in terms of the band structure.
Although there is a sophisticated approach to compute $D$\cite{freimuth2014}, we employ the current simple approach to explore various parameters and materials.
Furthermore, this approach is appropriate to obtain a guiding principle for controlling $D$ as discussed below.\\

\begin{figure}
	\begin{center}
	\includegraphics[scale=0.6]{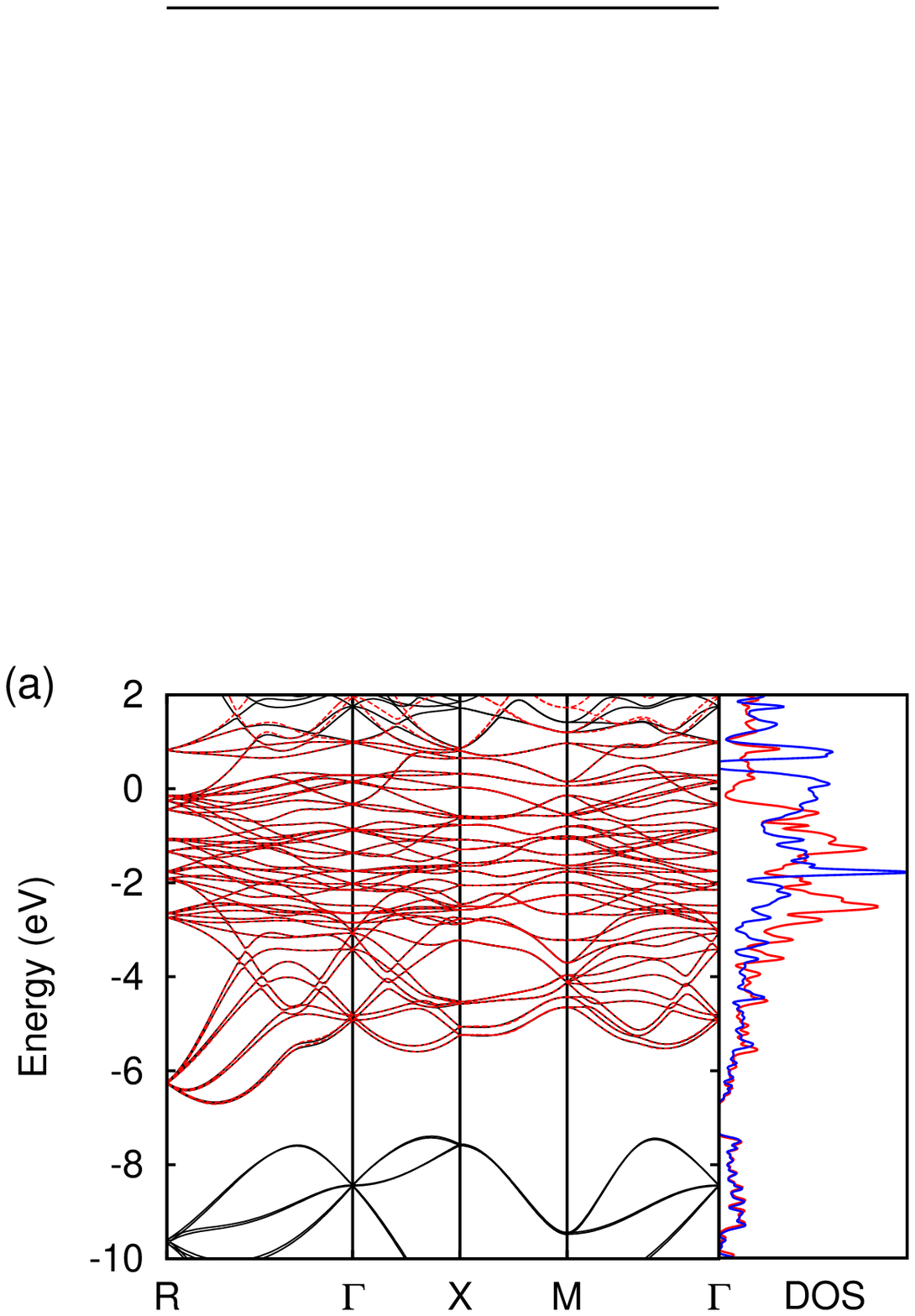}
	\includegraphics[scale=0.6]{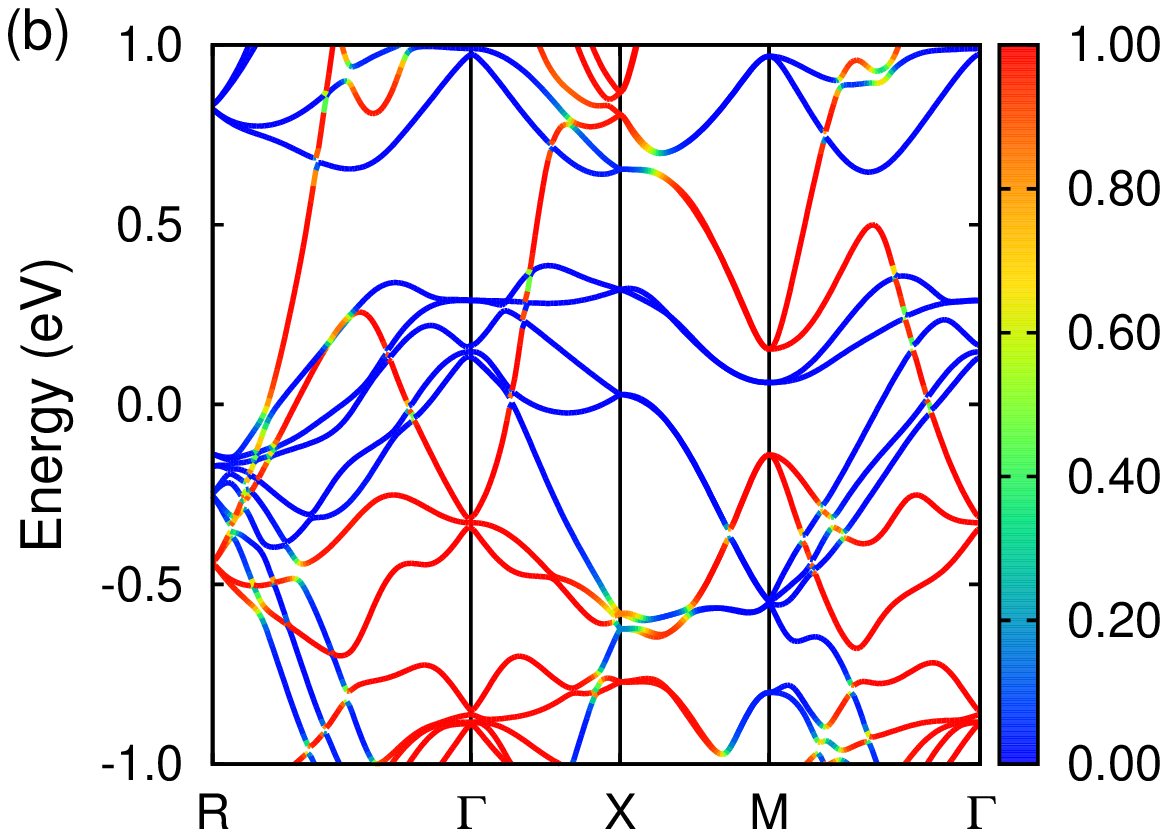}
	\caption{\textbf{Band structure of FeGe.} (a) Comparison between DFT band structure (black solid lines) and tight-binding band structure (red broken lines).
	Densities of states for up spin (red line) and down spin (blue line) are also shown.
	(b) Detailed band structure around the Fermi level with colors representing the weight of up spin; that is, red (blue) lines correspond to up-spin (down-spin) bands. The Fermi level is set to zero.}
	\label{fig:band}
	\end{center}
\end{figure}

\noindent\textbf{{\textit{Ab initio} band structure.}}
Figure \ref{fig:band}(a) shows the DFT band structure of FeGe (black solid lines).
Here, we include the spin-orbit couplings and assume the ferromagnetic moment along the $z$ axis.
The calculated local magnetic moment is 1.18 $\mu_B$ per Fe atom, which is consistent with the experiments\cite{wappling1968,lundgren1968} and previous calculations\cite{yamada2003}.
Using this electronic structure, we construct the tight-binding model made of Fe 3d and Ge 4p Wannier orbitals to reproduce the band structure below the Fermi level as shown in red broken lines.
The densities of states for up spin (red line) and down spin (blue line) are also shown in Fig.~\ref{fig:band}(a).
As can be seen, there is a large exchange splitting, $\Delta$.
According to the energy difference of up and down spins for the Fe 3d orbitals, we obtain $\Delta = 1.17$ eV.
In Fig.~\ref{fig:band}(b), the obtained tight-binding band structure around the Fermi level is illustrated with colors representing the weight of the up spin.
Since we consider ferromagnetic electronic structure, each band can be basically characterized as either up-spin or down-spin band as shown in Fig.~\ref{fig:band}(b).
In addition, due to the spin-orbit couplings, there are several anticrossing points where complex spin texture emerges.\\

\begin{figure}
	\includegraphics[scale=0.35]{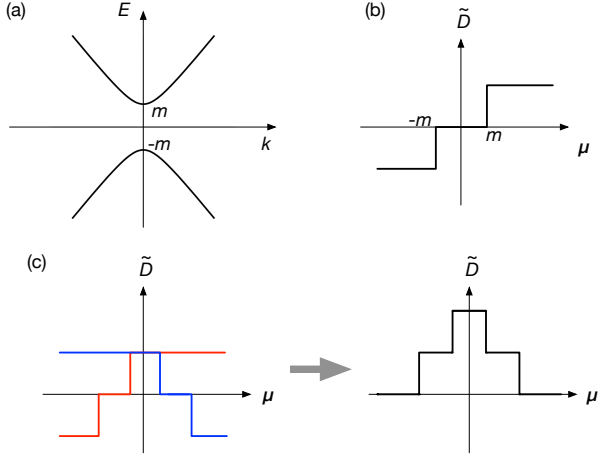}
	\caption{\textbf{Schematic pictures of the DM interaction around the band anticrossing points.}
	(a) Band structure of the two-band model defined in Eq. \eqref{eq:twoband} and (b) the chemical potential dependence of the DM interaction corresponding to this band structure.
    When two anticrossing points with different energies and spin textures with opposite chiralities reside close to each other, the peak structure appears as shown in (c).
	}
	\label{fig:schematic}
\end{figure}

\begin{figure}
	\includegraphics[scale=0.7]{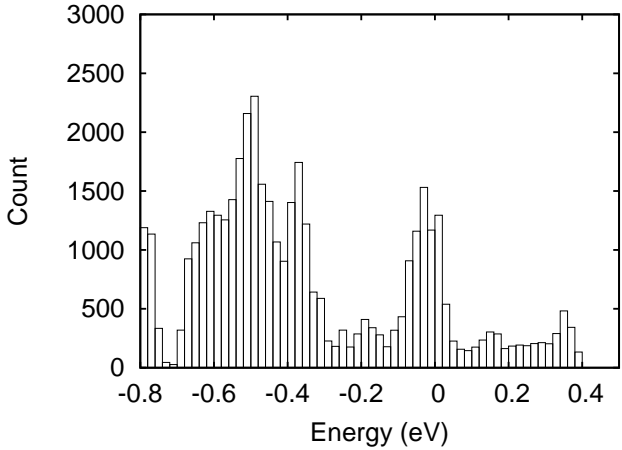}
	\caption{\textbf{Distribution of anticrossing points.}
	Number of k points in 64$\times$64$\times$64 mesh where the up-spin weight, $w_{\uparrow}$, satisfies $0.4 < w_{\uparrow} < 0.6$.
	}
	\label{fig:mix_point}
\end{figure}

\noindent\textbf{{\textit{Ab initio} evaluation of the DM interaction.}}
Let us start with the simple two-band model in two dimensions considered in Ref.~\onlinecite{wakatsuki2014}.
The Hamiltonian is represented by a $2\times2$ matrix,
\begin{align}
	H = k_x \sigma^x + k_y \sigma^y + m \sigma^z.
	\label{eq:twoband}
\end{align}
We assume that the band dispersion is
linear in the $k_x$-$k_y$ plane, and dispersionless in the $k_z$ direction.
We introduce $m$ to open a gap at the band crossing point as shown in Fig.~\ref{fig:schematic} (a). 
The static spin susceptibility can be calculated analytically, and the result
is 
% NN note chi^ab = Tr sigma^a G(k+q) sigma G(k)
\begin{align}
	\tilde{D}_y = \lim_{q \to 0} \frac{\partial \chi^{xz}}{\partial (i q^y)}
	&= \frac{1}{V} \sum_{\bm k} \frac{\delta(\mu-E_{\bm k}) - \delta(\mu + E_{\bm k})}{4 E_{\bm k}}\nonumber\\
	&= \frac{1}{8\pi} \left[ \theta(\mu - m) - \theta(-\mu - m) \right]
\end{align}
where $\mu$ is the chemical potential and $E_{\bm k} = \sqrt{k_x^2+k_y^2+m^2}$. 
It is interesting to note that $\tilde{D}_y$ is negative for $\mu<-m$, and positive for $\mu>m$ as shown Fig.~\ref{fig:schematic} (b).
If we reverse the spin texture by modifying the Hamiltonian as $\sigma_y \rightarrow -\sigma_y$, 
$\tilde{D}_y$ becomes positive for $\mu<-m$, and negative for $\mu>m$. 
This result suggests that the position of the Fermi level and the spin texture around the anticrossing point are crucial to determine the sign of $D$.

In real materials, the situation is not so simple as that of this two-band model.
The anticrossing points form complex surfaces in a four-dimensional space spanned by the
energy and the wave number, and the electronic states around neighboring 
anticrossing points can hybridize with each other. In Fig.~\ref{fig:schematic} (c), as a representative case, 
we schematically show how $\tilde{D}$ changes as a function of the chemical potential when
two anticrossing points with different energies and spin textures with opposite chiralities
reside close to each other. 
Thus, when many anticrossings are densely clustered around the Fermi level, the sign and the size of $\tilde{D}$ should change drastically. 
In fact, the band structure of FeGe has many anticrossing points around the Fermi level (see Fig.~\ref{fig:band}). 
To visualize the distribution of the anticrossing points as a function of energy, in Fig.~\ref{fig:mix_point}, we plot the number of $k$ points in $64\times64\times64$ mesh where the up-spin weight, $w_{\uparrow}$, satisfies $0.4 <w_{\uparrow}<0.6$.
We can see that there are two peaks around $\mu=0$ eV and $-0.5$ eV, where $\tilde{D}$ is expected to change significantly.

\begin{figure}
	\begin{center}
	\includegraphics[scale=0.45]{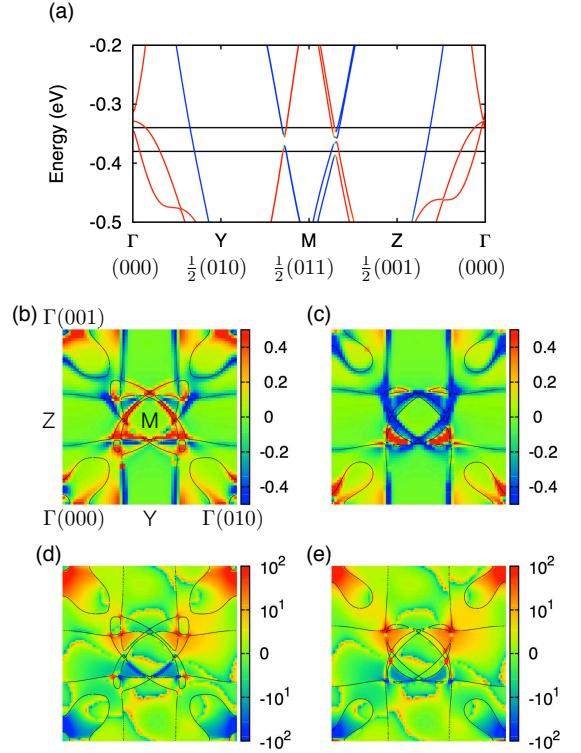}
	\caption{\textbf{Contribution of anticrossing points to the DM interaction.}
	(a) Band structure around the anticrossing points and $\bm k$-dependence of $\tilde{D}_x(\bm k)$ at two different chemical potentials (b) $\mu = -0.38$ eV and (c) $\mu = -0.34$ eV at $T=100$K. Fermi surface is also shown in black lines. For comparison, the Berry curvatures at (d) $\mu = -0.38$ eV and (e) $\mu = -0.34$ eV are plotted.}
	\label{fig:texture}
	\end{center}
\end{figure}

To analyze the contribution of each anticrossing point, next, we focus on the anticrossing points around the $M$ (0 1/2 1/2) point.
In Fig.~\ref{fig:texture}(a), there are two anticrossing points between Y and M points, and between M and Z points. %along the Y-M line and M-Z line.
In $k_x=0$ plane, such anticrossing points continuously form a closed loop around the M point.
In Fig.~\ref{fig:texture}(b) and (c), we show the $\bm k$-dependence of $\tilde{D}_x(\bm k)$ when the chemical potential is below and above the anticrossing points shown in black lines in Fig.~\ref{fig:texture}(a), respectively.
We see that the texture around the M point drastically changes when the chemical potential sweeps across the anticrossing points, which is consistent with the simple two-band calculation\cite{wakatsuki2014}.
Note that $\tilde{D}_x(\bm k)$ around the other Fermi surfaces such as the one between $\Gamma$-Y line are not negligible although the spin mixture is not so significant.
This is because the effect of spin mixing extends away from anticrossing points;
that is, $\tilde{D}(\bm k)$ has non-negligible values typically up to 0.2 - 0.3 eV away from anticrossing points.
For comparison, in Fig.~\ref{fig:texture} (d) and (e), we show the Berry curvature, $\Omega^z(\bm k)$, which is the origin of intrinsic anomalous Hall conductivity (AHC)\cite{fang2003,nagaosa2010}.
$\Omega^z(\bm k)$ is defined as
\begin{align}
	\Omega^z(\bm k) = - \hbar^2 \sum_{n\neq n'} f(\varepsilon_{n \bm k})
	\frac{2 {\rm Im} \langle n\bm k | v_x | n' \bm k \rangle \langle n'\bm k | v_y | n  \bm k \rangle}
	{ (\varepsilon_{n' \bm k} - \varepsilon_{n \bm k})^2 },
	\label{eq:berry}
\end{align}
where $v_x$ and $v_y$ are velocity operators.
Since the spin mixing is important for both cases, there are common regions where the contributions to the DM interaction and AHC are large.
However, in the Berry curvature, the summation in Eq.~\eqref{eq:berry} is restricted to $n \neq n'$ while it is not in Eq.~\eqref{eq:chiq}.
As a result, only the restricted region is important for AHC, which is in sharp contrast to $\tilde{D}$.

\begin{figure}
	\begin{center}
	\includegraphics[scale=0.7]{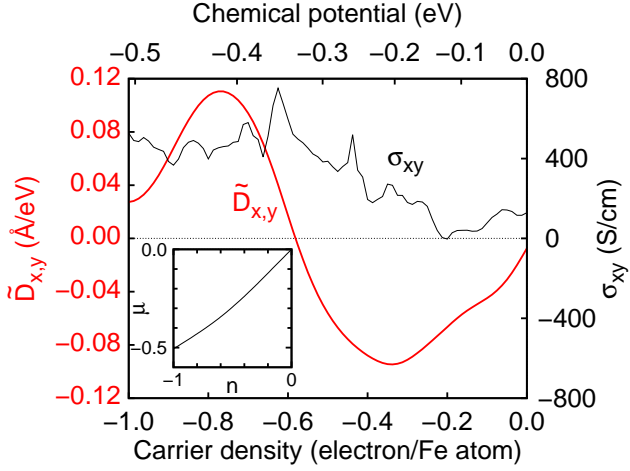}
	\caption{\textbf{Carrier density dependence of the DM interaction.}
	DM interaction coefficients, $\tilde{D}_x$, and $\tilde{D}_y$, and the AHC, $\sigma_{xy}$ as a function of the carrier density, $n$. The inset shows the relation between the chemical potential $\mu$ and carrier density $n$.
We use the rigid band approximation starting from the electronic structure of FeGe ($n=0.0$, $\mu=0.0$ eV). $n = -1.0$ ($\mu = -0.506$ eV) corresponds to the carrier density of MnGe.}
	\label{fig:ndep}
	\end{center}
\end{figure}

Figure~\ref{fig:ndep} shows the resulting $\tilde{D}_{x,y}$ at $T=300$K as a function of the carrier density together with the AHC, $\sigma_{xy}$.
In the calculation, we use the rigid band approximation.
The relation between the carrier density, $n$, and the chemical potential, $\mu$, is shown in the inset of Fig.~\ref{fig:ndep}.
At $\mu = -0.506$ eV, the number of hole is 1.0 per Fe atom, which corresponds to the carrier density in MnGe.
In Fig.~\ref{fig:ndep}, we find that $\tilde{D}_{x,y}$ shows clear sign change from FeGe $(\tilde{D}_{x,y} < 0)$ to MnGe $(\tilde{D}_{x,y} > 0)$,
which is consistent with the experimental sign change of skyrmion helicity\cite{shibata2013,grigoriev2013},
and is in sharp contrast to $\sigma_{xy}$.
The positive (negative) hump structure in $\tilde{D}(\mu)$ around $\mu \sim 0.4$ ($0.2$) eV originates from the peak structure around $\mu \sim 0.5$ ($0.0$) eV in Fig.~\ref{fig:mix_point}, respectively.
If we assume that contributions from $\mu\sim0(-0.5)$ eV is negative (positive), we can understand why $\tilde{D}$ changes its sign around $\mu\sim0.3$ eV.
As for $\sigma_{xy}$, the calculated value of $\sigma_{xy}$ in MnGe is larger than that in FeGe and there is no sign change.
This behavior including size and sign agrees well with the experimental anomalous Hall contribution to $\sigma_{xy}$ for MnGe\cite{kanazawa2011} and for \MnFeGe\cite{kanazawa2014}.\\

\begin{figure*}
	\begin{center}
	\includegraphics[scale=0.7]{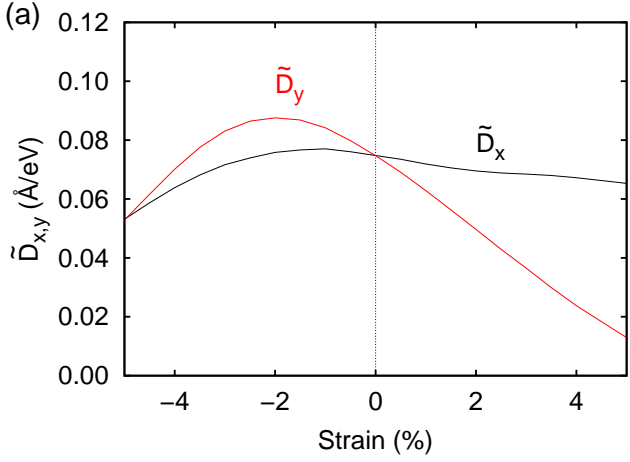}
	\includegraphics[scale=0.7]{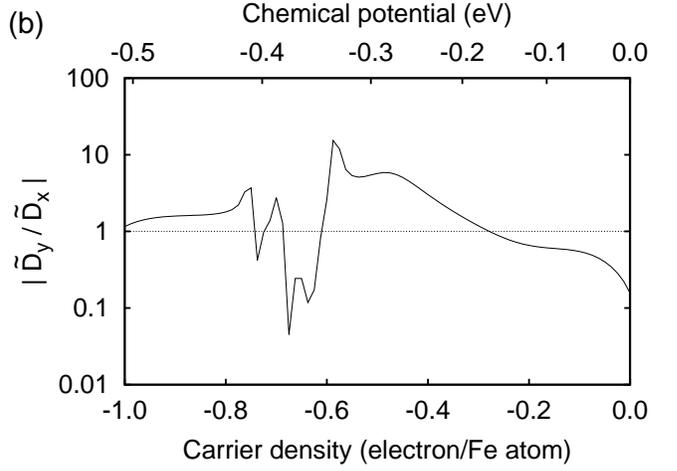}
	\caption{
		\textbf{Strain dependence of the DM interaction.}
		(a) $\tilde{D}_{x,y}$ at $n = -0.45$ as a function of uniaxial strain along the $y$ direction.
		(b) Anisotropy of the DM interaction, $|\tilde{D}_{y}/\tilde{D}_x|$ at +5\% strain along the $y$ direction as a function of the carrier density.
	}
	\label{fig:strain}
	\end{center}
\end{figure*}

\begin{figure*}
	\begin{center}
	\includegraphics[scale=0.7]{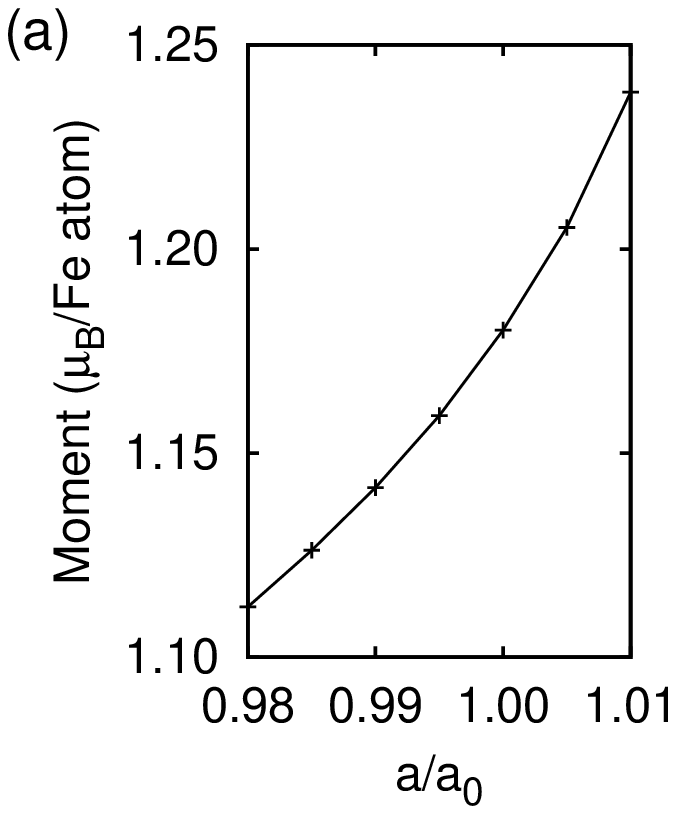}
	\includegraphics[scale=0.7]{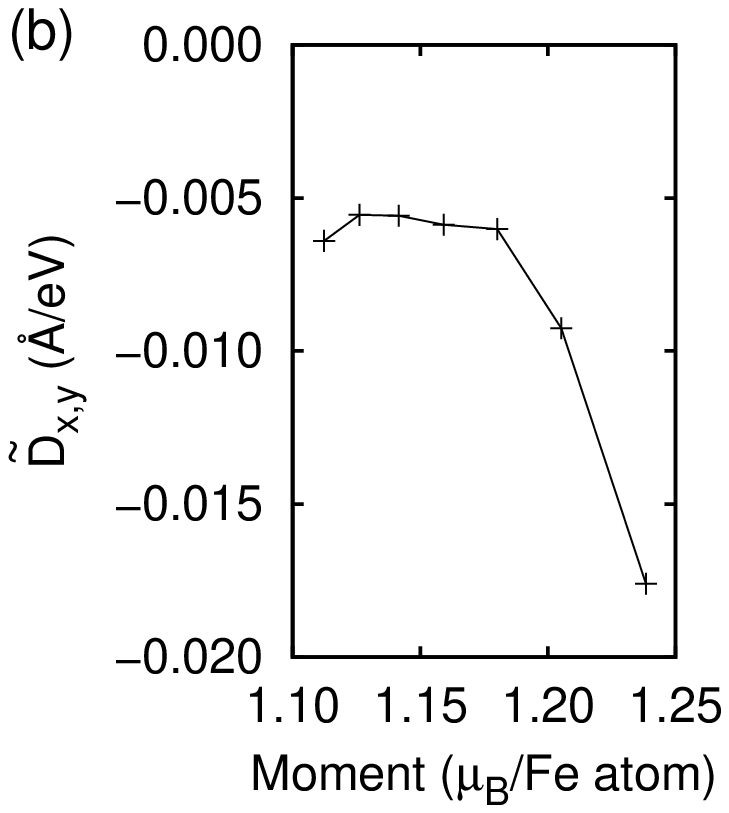}
	\includegraphics[scale=0.7]{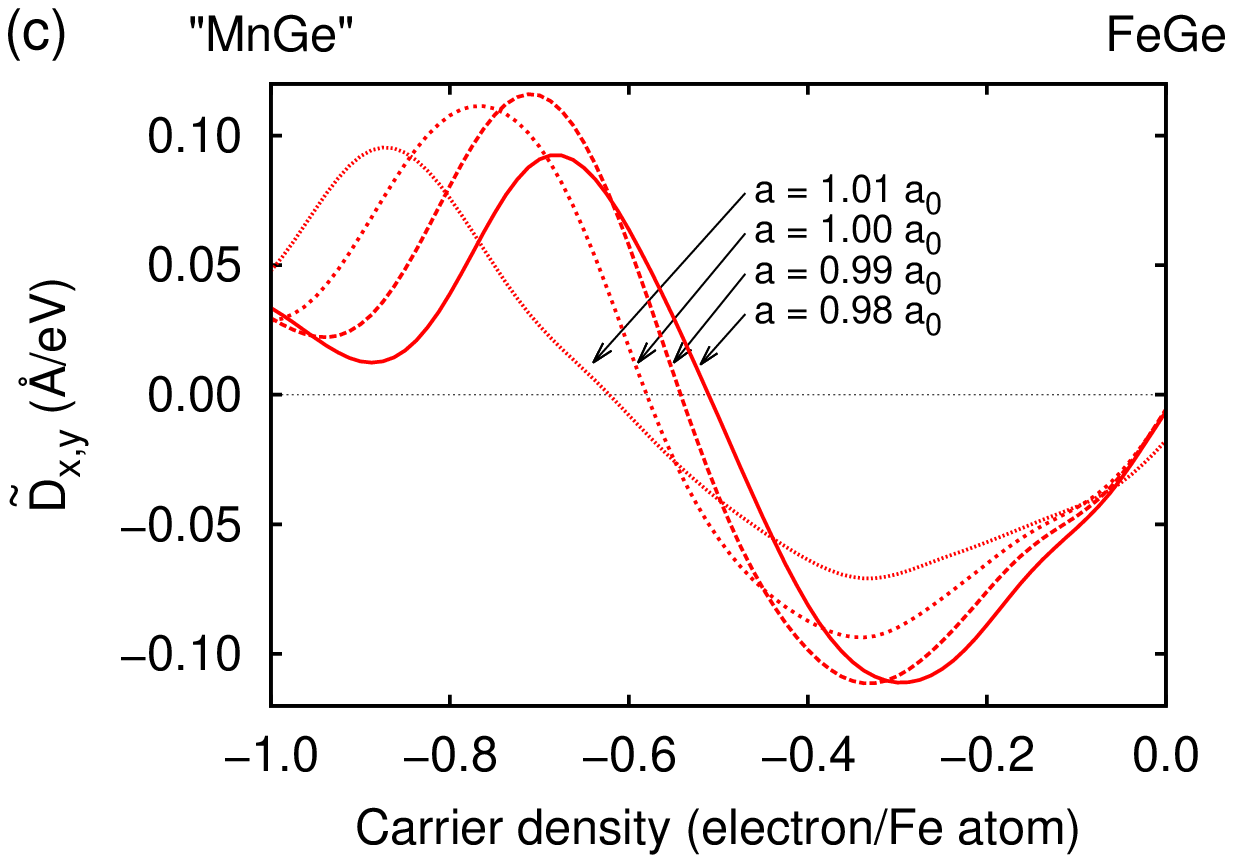}

	\includegraphics[scale=0.65]{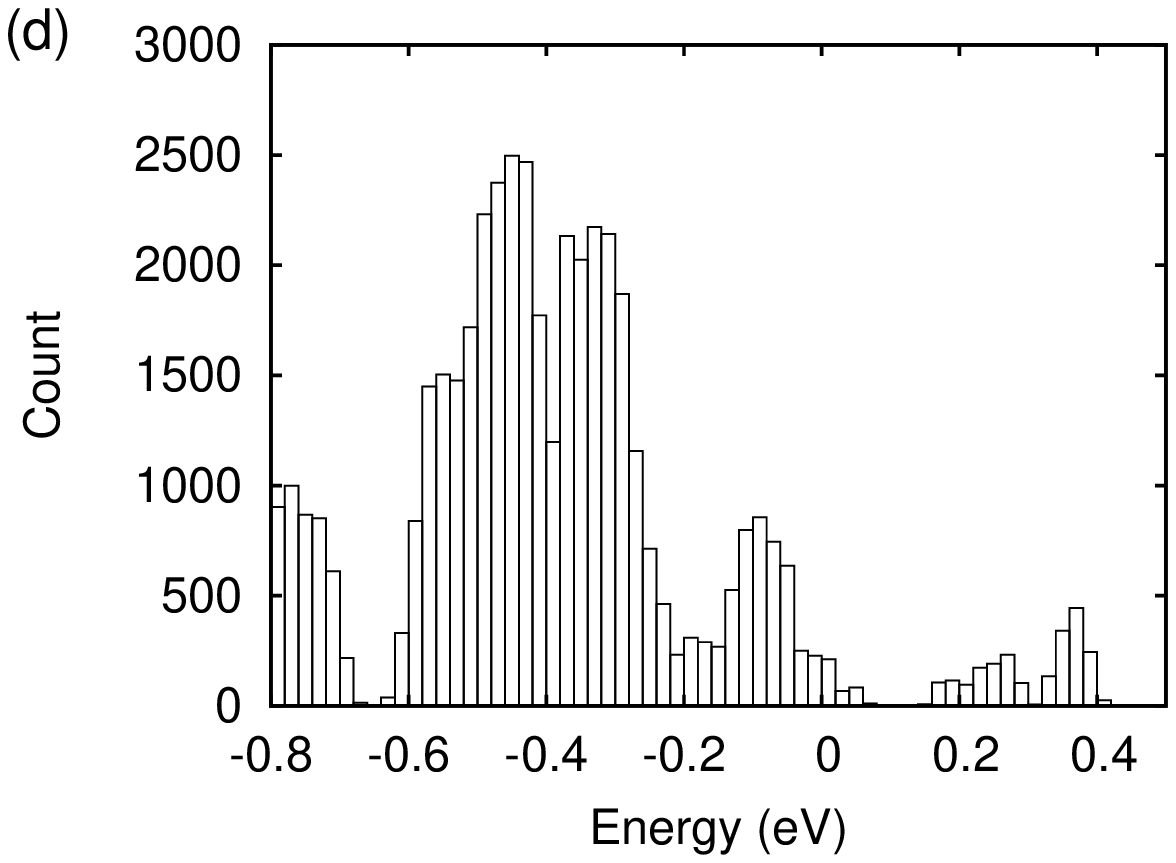}
	\includegraphics[scale=0.65]{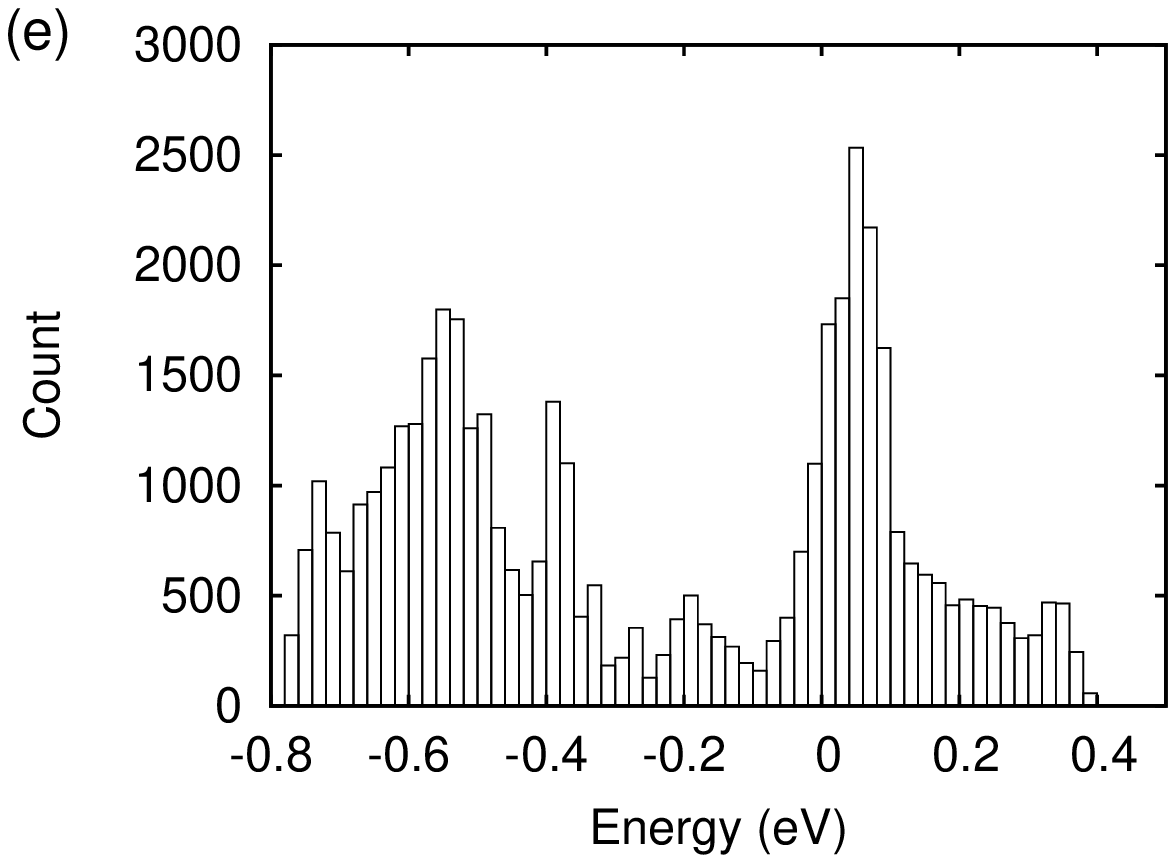}
	\caption{\textbf{Lattice constant dependence of the DM interaction.}
	(a) Lattice constant dependence of the magnetic moment on each Fe atom for FeGe and (b) corresponding magnetic moment dependence of $\tilde{D}_{x,y}$.
	(c) $\tilde{D}_{x,y}$ with several lattice constants, $a/a_0 = 0.98, 0.99, 1.00$, and $1.01$.
	Distribution of anticrossing points at (d) $a/a_0 = 0.99$, and (e) $a/a_0 = 1.01$.
}
	\label{fig:pdep}
	\end{center}
\end{figure*}

\noindent\textbf{{Strain-induced huge anisotropy of the DM interaction.}}
According to the above discussion, there are several ways to change the size and sign of $D$.
As we have seen above, $D$ can be efficiently controlled by carrier doping. 
We can also exploit the temperature dependence of the exchange splitting.
When the relative position of up- and down- spin band changes, the distribution of anticrossing points in the band structure also changes, which will have a direct impact on $D$.
%In fact, the dependence of $D$ on the moment is an important issue as observed in MnGe
As an example, we will show later the change in $\tilde{D}_{x,y}$ of FeGe as a function of the moment per Fe atom in Fig.~\ref{fig:pdep} (b).
This mechanism can be related to the temperature dependence of the magnetic moment and skyrmion size in MnGe\cite{kanazawa2011}.
Another interesting possibility is to make use of the strain effect.
If we apply a strain to the system, the symmetry of the electronic structure can be lowered, and the distribution of the anticrossing points will change drastically. 
This effect is expected to be prominent especially when $D$ changes its sign. 
Figure \ref{fig:strain} (a) shows the calculated $\tilde{D}_{x,y}$ at $n = -0.45$ for which we apply the uniaxial strain along the $y$ direction.
We find that the difference between $\tilde{D}_x$ and $\tilde{D}_y$ actually enhances particularly by the elongation along the $y$ axis; $\tilde{D}_x$ is about 40 \% (400 \%) larger than $\tilde{D}_y$ at +2 \% (+5\%) strain.
The carrier density dependence of the anisotropy, $|\tilde{D}_{y}/\tilde{D}_{x}|$, for fixed strain of +5\% is shown in Fig.~\ref{fig:strain} (b).
We can see that the anisotropy becomes large particularly around the region of sign change.\\

\begin{figure}
	\begin{center}
	\includegraphics[scale=0.7]{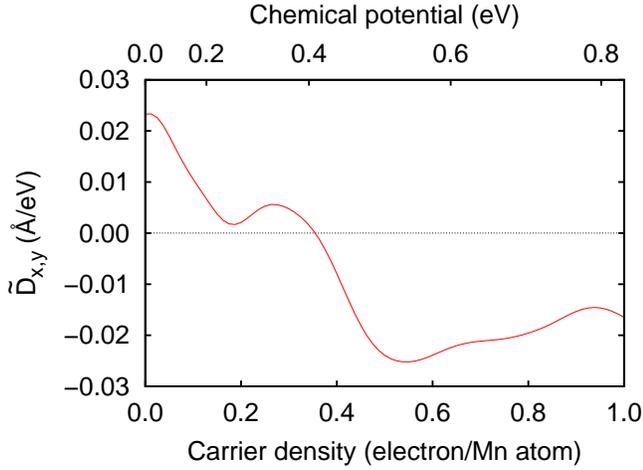}
	\caption{\textbf{Carrier density dependence of the DM interaction starting from MnGe electronic structure.}
		Carrier density dependence of the DM interaction coefficients, $\tilde{D}_x$, and $\tilde{D}_y$.
		We use the electronic structure of MnGe ($n=0.0$) for the same crystal structure as that of FeGe and employ the rigid band approximation.
		$n=1.0$ ($\mu = 0.832$ eV) corresponds to the carrier density of FeGe.
}
	\label{fig:MnGe}
	\end{center}
\end{figure}

%\section{Discussion}
\noindent\textbf{\large{Discussion}}\\
In the present calculation, we employed the rigid band approximation.
To examine its validity, we have performed a calculation for the other end material MnGe and doped negative carriers by the rigid band approximation.
As shown in Fig.~\ref{fig:MnGe}, we have obtained a qualitatively similar result in that $\tilde{D}$ is positive (negative) for the end material MnGe (FeGe).
Thus the result that $\tilde{D}$ for \MnFeGe\ changes its sign between $x=0$ and $1$ should be robust, even when we go beyond the rigid band approximation.

Regarding the crystal structure, it has been known that the magnetic moment for the optimized structure is much smaller than the experimental value within the local density approximation\cite{neef2009}.
On the other hand, for the experimental structure, the size of magnetic moment is similar to that in the experiment ($\sim 1\mu_{\rm B}$).
In fact, within our calculations, the magnetic moment in FeGe decreases with decreasing the lattice constant as shown in Fig.~\ref{fig:pdep} (a), which is consistent with previous studies\cite{yamada2003,jarlborg2004}.
As a result, $\tilde{D}$ exhibits non-trivial magnetic moment dependence as shown in Fig.~\ref{fig:pdep} (b).
Since the magnetic moment dependence of $D$ is also an important issue, let us next discuss how the change in the lattice structure or the magnetic moment affect $D$ in \MnFeGe.
In Fig.~\ref{fig:pdep} (c), we plot the lattice constant dependence of $\tilde{D}$, together with the distribution of band anti-crossings (d), (e) for lattice constants $a=0.99a_0$ and $1.01a_0$, where $a_0$ is the experimental lattice constant.
As can be seen, for larger $a$ and magnetic moment, the energy difference of two peaks in the histogram of Fig.~\ref{fig:mix_point} becomes larger (Fig.~\ref{fig:pdep}(e)).
Consequently, the density at which $\tilde{D}$ changes its sign ($n_c$) becomes larger (Fig.~\ref{fig:pdep}(c)).
However, the qualitative feature of $\tilde{D}$ is robust against the change in $a$.
Therefore, our calculation successfully explains why $\tilde{D}$ is positive (negative) for MnGe (FeGe), and provides a useful guideline for materials design of skyrmion crystal in metallic helimagnets.

\begin{table}
	\caption{\textbf{List of parameters for FeGe and MnGe.}
	Transition temperatures, $T_N$ (K), ferromagnetic exchange couplings, $J$ (meV\AA$^2$), helical periods, $\lambda$(\AA), experimental DM interactions evaluated by $D = 4\pi J/\lambda$ (meV\AA) and calculated $D$ (meV\AA).
	$J$ is evaluated by assuming $J \propto T_N$ and the values of $J=52$ meV\AA$^2$ and $T_N = 30$ K for MnSi.
}
	\begin{tabular}{cccccc} 
		\hline
		\hline
		& $T_N$ & $J$ & $\lambda$ & $D$ (expt.) & $D$ (calc.) \\
		& (K) & (meV\AA$^2$) & (\AA) & (meV\AA) & (meV\AA) \\
		\hline
		FeGe & 278 & 482 & 700 & -8.7 & -10.1\\
		MnGe & 170 & 295 &  30 &  124 &  107\\
		\hline
		\hline
	\end{tabular}
	\label{tab:parameters}
\end{table}
% FeGe: TN = 278, lambda = 70  \tilde{D} = 0.0074
% MnGe: TN = 170, lambda = 3   \tilde{D} = 0.0233
% MnSi: TN = 30,  lambda = 18
Finally, let us compare the quantitative values of $\tilde{D}$ with experiments.
For this purpose, we should rescale $\tilde{D}$ using the exchange splitting as $D = \Delta^2 \tilde{D}$\cite{wakatsuki2014}.
Using $\Delta = 1.17$ eV for FeGe and $\Delta = 2.14$ eV for MnGe based on the calculation shown in Fig.~\ref{fig:MnGe}, we can obtain $D = - 10.1$ meV\AA\ for FeGe and $D = 107$ meV\AA\ for MnGe.
The experimental values of $D$ can be estimated using $J$ and the helical period, $\lambda$ as $D = 4 \pi J/\lambda$.
Assuming that $J$ scales to $T_N$ and using the values of $J=52$ meV\AA$^2$ and $T_N = 30$ K for MnSi\cite{ishikawa1977,freimuth2013}, the experimental values of $D$ for FeGe and MnGe are -8.7 meV\AA\ and 124 meV\AA, respectively, which are in good agreement with our results (see Table I).
\\

%\section{Method}
\noindent\textbf{\large{Method}}\\
%\subsection{Evaluation of DM interaction}
\noindent\textbf{{Crystal structure.}}
In the calculations, experimental values are used for the crystallographic parameters\cite{lebech1989}.
For the pressure and strain calculations, internal coordinates of the atoms are fixed and only the lattice parameters are modified.
To symmetrize $\tilde{D}_x$ and $\tilde{D}_y$, we use two different internal coordinates of atoms in the (4a) position, that is, $(x,x,x)$ where $x_{\rm Fe,Mn} = 0.135$ and $x_{\rm Ge} = 0.842$, and the one with its 90-degree rotation along the $z$-axis and take the average.
Note that our structure is right-handed according to Ref.~\onlinecite{shibata2013}.
In the experiment, observed skyrmions on the right-handed crystal structure in FeGe (MnGe) are anticlockwise (clockwise), indicating that $D_{x,y} < 0$ ($D_{x,y} > 0$).\\

\noindent\textbf{{DFT calculations.}}
To evaluate $\chi_0^{\alpha \beta}$ in FeGe, we perform the electronic structure calculation within the generalized-gradient approximation (GGA)\cite{pbe1996} based on the density functional theory\cite{espresso}.
We use ultrasoft pseudopotentials\cite{vanderbilt1990} and plain-wave basis sets to describe the charge densities and wave functions with cutoff energies of 40Ry and 500Ry, respectively.
We use $8 \times 8 \times 8$ $k$-point mesh.
With including the spin-orbit couplings and assuming the ferromagnetic moment along the $z$ axis, we obtain non-collinear magnetic structure.
Using this electronic structure, we calculate Wannier functions for Fe 3d and Ge 4p orbitals using wannier90 code\cite{marzari1997,souza2001,mostofi2008}.
Based on the Wannier functions, we construct a tight-binding model on the restricted Hilbert space and calculate $\chi^{\alpha \beta}$ using $64 \times 64 \times 64$ k-point mesh at $T=300$K.
The AHC is also calculated using the Wannier interpolation technique with $200 \times 200 \times 200$ k-point mesh\cite{wang2006}.\\

%\begin{acknowledgments}
\noindent\textbf{\large{Acknowledgements}}\\
The authors thank N. Kanazawa, W. Koshibae, D. Morikawa, K. Shibata, and Y. Tokura for helpful discussions.
This work is supported by Grant-in-Aids for Scientific Research (No. 24224009 and No. 25104711) from the Ministry of Education, Culture, Sports, Science and Technology (MEXT) of Japan, and ImPACT Program of Council for Science, Technology and Innovation (Cabinet Office, Government of Japan).\\
%\end{acknowledgments}

\noindent\textbf{\large{Author contributions}}\\
T.K. carried out the numerical calculations. T.K., N.N. and R.A. analyzed the results and wrote the paper.


\begin{thebibliography}{10}
\expandafter\ifx\csname url\endcsname\relax
  \def\url#1{\texttt{#1}}\fi
\expandafter\ifx\csname urlprefix\endcsname\relax\def\urlprefix{URL }\fi
\providecommand{\bibinfo}[2]{#2}
\providecommand{\eprint}[2][]{\url{#2}}

\bibitem{bogdanov1989}
\bibinfo{author}{Bogdanov, A.~N.} \& \bibinfo{author}{Yablonskii, D.~A.}
\newblock \bibinfo{title}{{Thermodynamically stable ``vortices'' in
  magnetically ordered crystals. The mixed state of magnets}}.
\newblock \emph{\bibinfo{journal}{Sov. Phys. JETP}}
  \textbf{\bibinfo{volume}{68}}, \bibinfo{pages}{101--103}
  (\bibinfo{year}{1989}).

\bibitem{rossler2006}
\bibinfo{author}{R{\"o}{\ss}ler, U.~K.}, \bibinfo{author}{Bogdanov, A.~N.} \&
  \bibinfo{author}{Pfleiderer, C.}
\newblock \bibinfo{title}{{Spontaneous skyrmion ground states in magnetic
  metals}}.
\newblock \emph{\bibinfo{journal}{Nature}} \textbf{\bibinfo{volume}{442}},
  \bibinfo{pages}{797--801} (\bibinfo{year}{2006}).

\bibitem{muhlbauer2009}
\bibinfo{author}{M{\"u}hlbauer, S.} \emph{et~al.}
\newblock \bibinfo{title}{{Skyrmion Lattice in a Chiral Magnet}}.
\newblock \emph{\bibinfo{journal}{Science}} \textbf{\bibinfo{volume}{323}},
  \bibinfo{pages}{915--919} (\bibinfo{year}{2009}).

\bibitem{yu2010}
\bibinfo{author}{Yu, X.~Z.} \emph{et~al.}
\newblock \bibinfo{title}{{Real-space observation of a two-dimensional skyrmion
  crystal}}.
\newblock \emph{\bibinfo{journal}{Nature}} \textbf{\bibinfo{volume}{465}},
  \bibinfo{pages}{901--904} (\bibinfo{year}{2010}).

\bibitem{yu2011}
\bibinfo{author}{Yu, X.~Z.} \emph{et~al.}
\newblock \bibinfo{title}{{Near room-temperature formation of a skyrmion
  crystal in thin-films of the helimagnet FeGe}}.
\newblock \emph{\bibinfo{journal}{Nature Materials}}
  \textbf{\bibinfo{volume}{10}}, \bibinfo{pages}{106--109}
  (\bibinfo{year}{2011}).

\bibitem{nagaosa2013}
\bibinfo{author}{Nagaosa, N.} \& \bibinfo{author}{Tokura, Y.}
\newblock \bibinfo{title}{{Topological properties and dynamics of magnetic
  skyrmions}}.
\newblock \emph{\bibinfo{journal}{Nature Nanotech.}}
  \textbf{\bibinfo{volume}{8}}, \bibinfo{pages}{899--911}
  (\bibinfo{year}{2013}).

\bibitem{shibata2013}
\bibinfo{author}{Shibata, K.} \emph{et~al.}
\newblock \bibinfo{title}{{Towards control of the size and helicity of
  skyrmions in helimagnetic alloys by spin-orbit coupling}}.
\newblock \emph{\bibinfo{journal}{Nature Nanotech.}}
  \textbf{\bibinfo{volume}{8}}, \bibinfo{pages}{723--728}
  (\bibinfo{year}{2013}).

\bibitem{grigoriev2013}
\bibinfo{author}{Grigoriev, S.~V.} \emph{et~al.}
\newblock \bibinfo{title}{{Chiral Properties of Structure and Magnetism in
  Mn$_{1-x}$Fe$_{x}$Ge Compounds: When the Left and the Right are Fighting, Who
  Wins? }}.
\newblock \emph{\bibinfo{journal}{Phys. Rev. Lett.}}
  \textbf{\bibinfo{volume}{110}}, \bibinfo{pages}{207201}
  (\bibinfo{year}{2013}).

\bibitem{seki2012}
\bibinfo{author}{Seki, S.}, \bibinfo{author}{Yu, X.~Z.},
  \bibinfo{author}{Ishiwata, S.} \& \bibinfo{author}{Tokura, Y.}
\newblock \bibinfo{title}{{Observation of Skyrmions in a Multiferroic
  Material}}.
\newblock \emph{\bibinfo{journal}{Science}} \textbf{\bibinfo{volume}{336}},
  \bibinfo{pages}{198} (\bibinfo{year}{2012}).

\bibitem{janson2014}
\bibinfo{author}{Janson, O.} \emph{et~al.}
\newblock \bibinfo{title}{{The quantum nature of skyrmions and half-skyrmions
  in Cu2OSeO3}}.
\newblock \emph{\bibinfo{journal}{Nature Communications}}
  \textbf{\bibinfo{volume}{5}}, \bibinfo{pages}{1--11} (\bibinfo{year}{2014}).

\bibitem{heide2008}
\bibinfo{author}{Heide, M.}, \bibinfo{author}{Bihlmayer, G.} \&
  \bibinfo{author}{Bl{\"u}gel, S.}
\newblock \bibinfo{title}{{Dzyaloshinskii-Moriya interaction accounting for the
  orientation of magnetic domains in ultrathin films: Fe/W(110)}}.
\newblock \emph{\bibinfo{journal}{Phys. Rev. B}} \textbf{\bibinfo{volume}{78}},
  \bibinfo{pages}{140403} (\bibinfo{year}{2008}).

\bibitem{ferriani2008}
\bibinfo{author}{Ferriani, P.} \emph{et~al.}
\newblock \bibinfo{title}{{Atomic-Scale Spin Spiral with a Unique Rotational
  Sense: Mn Monolayer on W(001)}}.
\newblock \emph{\bibinfo{journal}{Phys. Rev. Lett.}}
  \textbf{\bibinfo{volume}{101}}, \bibinfo{pages}{027201}
  (\bibinfo{year}{2008}).

\bibitem{heide2009}
\bibinfo{author}{Heide, M.}, \bibinfo{author}{Bihlmayer, G.} \&
  \bibinfo{author}{Bl{\"u}gel, S.}
\newblock \bibinfo{title}{{Describing Dzyaloshinskii{\textendash}Moriya spirals
  from first principles}}.
\newblock \emph{\bibinfo{journal}{Physica B: Phys. Cond. Matt.}}
  \textbf{\bibinfo{volume}{404}}, \bibinfo{pages}{2678--2683}
  (\bibinfo{year}{2009}).

\bibitem{udvardi2003}
\bibinfo{author}{Udvardi, L.}, \bibinfo{author}{Szunyogh, L.},
  \bibinfo{author}{Palot{\'a}s, K.} \& \bibinfo{author}{Weinberger, P.}
\newblock \bibinfo{title}{{First-principles relativistic study of spin waves in
  thin magnetic films}}.
\newblock \emph{\bibinfo{journal}{Phys. Rev. B}} \textbf{\bibinfo{volume}{68}},
  \bibinfo{pages}{104436} (\bibinfo{year}{2003}).

\bibitem{ebert2009}
\bibinfo{author}{Ebert, H.} \& \bibinfo{author}{Mankovsky, S.}
\newblock \bibinfo{title}{{Anisotropic exchange coupling in diluted magnetic
  semiconductors: Ab initio spin-density functional theory}}.
\newblock \emph{\bibinfo{journal}{Phys. Rev. B}} \textbf{\bibinfo{volume}{79}},
  \bibinfo{pages}{045209} (\bibinfo{year}{2009}).

\bibitem{katsnelson2010}
\bibinfo{author}{Katsnelson, M.~I.}, \bibinfo{author}{Kvashnin, Y.~O.},
  \bibinfo{author}{Mazurenko, V.~V.} \& \bibinfo{author}{Lichtenstein, A.~I.}
\newblock \bibinfo{title}{{Correlated band theory of spin and orbital
  contributions to Dzyaloshinskii-Moriya interactions}}.
\newblock \emph{\bibinfo{journal}{Phys. Rev. B}} \textbf{\bibinfo{volume}{82}},
  \bibinfo{pages}{100403} (\bibinfo{year}{2010}).

\bibitem{dmitrienko2014}
\bibinfo{author}{Dmitrienko, V.~E.} \emph{et~al.}
\newblock \bibinfo{title}{{Measuring the Dzyaloshinskii{\textendash}Moriya
  interaction in a weak ferromagnet}}.
\newblock \emph{\bibinfo{journal}{Nature Physics}}
  \textbf{\bibinfo{volume}{10}}, \bibinfo{pages}{202--206}
  (\bibinfo{year}{2014}).

\bibitem{freimuth2013}
\bibinfo{author}{Freimuth, F.}, \bibinfo{author}{Bamler, R.},
  \bibinfo{author}{Mokrousov, Y.} \& \bibinfo{author}{Rosch, A.}
\newblock \bibinfo{title}{{Phase-space Berry phases in chiral magnets:
  Dzyaloshinskii-Moriya interaction and the charge of skyrmions}}.
\newblock \emph{\bibinfo{journal}{Phys. Rev. B}} \textbf{\bibinfo{volume}{88}},
  \bibinfo{pages}{214409} (\bibinfo{year}{2013}).

\bibitem{freimuth2014}
\bibinfo{author}{Freimuth, F.}, \bibinfo{author}{Bl{\"u}gel, S.} \&
  \bibinfo{author}{Mokrousov, Y.}
\newblock \bibinfo{title}{{Berry phase theory of
  Dzyaloshinskii{\textendash}Moriya interaction and spin{\textendash}orbit
  torques}}.
\newblock \emph{\bibinfo{journal}{J. Phys.-Cond. Matt.}}
  \textbf{\bibinfo{volume}{26}}, \bibinfo{pages}{104202}
  (\bibinfo{year}{2014}).

\bibitem{wakatsuki2014}
\bibinfo{author}{Wakatsuki, R.}, \bibinfo{author}{Ezawa, M.} \&
  \bibinfo{author}{Nagaosa, N.}
\newblock \bibinfo{title}{Domain wall of a ferromagnet on a three-dimensional
  topological insulator}.
\newblock \bibinfo{note}{ArXiv:1412.7910}.

\bibitem{nagaosa2010}
\bibinfo{author}{Nagaosa, N.}, \bibinfo{author}{Sinova, J.},
  \bibinfo{author}{Onoda, S.}, \bibinfo{author}{MacDonald, A.~H.} \&
  \bibinfo{author}{Ong, N.~P.}
\newblock \bibinfo{title}{{Anomalous Hall effect}}.
\newblock \emph{\bibinfo{journal}{Rev. Mod. Phys.}}
  \textbf{\bibinfo{volume}{82}}, \bibinfo{pages}{1539--1592}
  (\bibinfo{year}{2010}).

\bibitem{wappling1968}
\bibinfo{author}{W\"appling, R.} \& \bibinfo{author}{H\"aggstr\"om, L.}
\newblock \bibinfo{title}{{Mossbauer study of cubic FeGe}}.
\newblock \emph{\bibinfo{journal}{Phys. Lett. A}}
  \textbf{\bibinfo{volume}{28A}}, \bibinfo{pages}{173} (\bibinfo{year}{1968}).

\bibitem{lundgren1968}
\bibinfo{author}{Lundgren, L.}, \bibinfo{author}{Blom, K.~A.} \&
  \bibinfo{author}{Beckman, O.}
\newblock \bibinfo{title}{{Magnetic susceptibility measurements on cubic
  FeGe}}.
\newblock \emph{\bibinfo{journal}{Phys. Lett. A}}
  \textbf{\bibinfo{volume}{28A}}, \bibinfo{pages}{175} (\bibinfo{year}{1968}).

\bibitem{yamada2003}
\bibinfo{author}{Yamada, H.}, \bibinfo{author}{Terao, K.},
  \bibinfo{author}{Ohta, H.} \& \bibinfo{author}{Kulatov, E.}
\newblock \bibinfo{title}{{Electronic structure and magnetism of FeGe with
  B20-type structure}}.
\newblock \emph{\bibinfo{journal}{Physica B}}
  \textbf{\bibinfo{volume}{329-333}}, \bibinfo{pages}{1131--1133}
  (\bibinfo{year}{2003}).

\bibitem{fang2003}
\bibinfo{author}{Fang, Z.} \emph{et~al.}
\newblock \bibinfo{title}{{The Anomalous Hall Effect and Magnetic Monopoles in
  Momentum Space}}.
\newblock \emph{\bibinfo{journal}{Science}} \textbf{\bibinfo{volume}{302}},
  \bibinfo{pages}{92--95} (\bibinfo{year}{2003}).

\bibitem{kanazawa2011}
\bibinfo{author}{Kanazawa, N.} \emph{et~al.}
\newblock \bibinfo{title}{{Large Topological Hall Effect in a Short-Period
  Helimagnet MnGe}}.
\newblock \emph{\bibinfo{journal}{Phys. Rev. Lett.}}
  \textbf{\bibinfo{volume}{106}}, \bibinfo{pages}{156603}
  (\bibinfo{year}{2011}).

\bibitem{kanazawa2014}
\bibinfo{author}{Kanazawa, N.}
\newblock \bibinfo{title}{{Charge and heat transport phenomena in electronic
  and spin structures in B20-type compounds}} (\bibinfo{year}{2014}).
\newblock \bibinfo{note}{PhD thesis, Univ. of Tokyo}.

\bibitem{neef2009}
\bibinfo{author}{Neef, M.}, \bibinfo{author}{Doll, K.} \&
  \bibinfo{author}{Zwicknagl, G.}
\newblock \bibinfo{title}{{Ab initio study of pressure-induced metal-insulator
  transition in cubic FeGe}}.
\newblock \emph{\bibinfo{journal}{Phys. Rev. B}} \textbf{\bibinfo{volume}{80}},
  \bibinfo{pages}{035122} (\bibinfo{year}{2009}).

\bibitem{jarlborg2004}
\bibinfo{author}{Jarlborg, T.}
\newblock \bibinfo{title}{{Electronic structure and magnetism for from
  supercell calculations}}.
\newblock \emph{\bibinfo{journal}{Journal of Magnetism and Magnetic Materials}}
  \textbf{\bibinfo{volume}{283}}, \bibinfo{pages}{238--246}
  (\bibinfo{year}{2004}).

\bibitem{ishikawa1977}
\bibinfo{author}{Ishikawa, Y.}, \bibinfo{author}{Shirane, G.},
  \bibinfo{author}{Tarvin, J.~A.} \& \bibinfo{author}{Kohgi, M.}
\newblock \bibinfo{title}{{Magnetic excitations in the weak itinerant
  ferromagnet MnSi}}.
\newblock \emph{\bibinfo{journal}{Phys. Rev. B}} \textbf{\bibinfo{volume}{16}},
  \bibinfo{pages}{4956} (\bibinfo{year}{1977}).

\bibitem{lebech1989}
\bibinfo{author}{Lebech, B.}, \bibinfo{author}{Bernhard, J.} \&
  \bibinfo{author}{Freltoft, T.}
\newblock \bibinfo{title}{{Magnetic structures of cubic FeGe studied by
  small-angle neutron scattering}}.
\newblock \emph{\bibinfo{journal}{J. Phys. Cond. Matt.}}
  \textbf{\bibinfo{volume}{1}}, \bibinfo{pages}{6105--6122}
  (\bibinfo{year}{1989}).

\bibitem{pbe1996}
\bibinfo{author}{Perdew, J.~P.}, \bibinfo{author}{Burke, K.} \&
  \bibinfo{author}{Ernzerhof, M.}
\newblock \bibinfo{title}{{Generalized Gradient Approximation Made Simple}}.
\newblock \emph{\bibinfo{journal}{Phys. Rev. Lett.}}
  \textbf{\bibinfo{volume}{77}}, \bibinfo{pages}{3865} (\bibinfo{year}{1996}).

\bibitem{espresso}
\bibinfo{author}{Giannozzi, P.} \emph{et~al.}
\newblock \bibinfo{title}{{QUANTUM ESPRESSO: a modular and open-source software
  project for quantum simulations of materials}}.
\newblock \emph{\bibinfo{journal}{J. Phys. Condens. Matter}}
  \textbf{\bibinfo{volume}{21}}, \bibinfo{pages}{395502}
  (\bibinfo{year}{2009}).

\bibitem{vanderbilt1990}
\bibinfo{author}{Vanderbilt, D.}
\newblock \bibinfo{title}{{Soft self-consistent pseudopotentials in a
  generalized eigenvalue formalism}}.
\newblock \emph{\bibinfo{journal}{Phys. Rev. B}} \textbf{\bibinfo{volume}{41}},
  \bibinfo{pages}{7892} (\bibinfo{year}{1990}).

\bibitem{marzari1997}
\bibinfo{author}{Marzari, N.} \& \bibinfo{author}{Vanderbilt, D.}
\newblock \bibinfo{title}{{Maximally localized generalized Wannier functions
  for composite energy bands}}.
\newblock \emph{\bibinfo{journal}{Phys. Rev. B}} \textbf{\bibinfo{volume}{56}},
  \bibinfo{pages}{12847--12865} (\bibinfo{year}{1997}).

\bibitem{souza2001}
\bibinfo{author}{Souza, I.}, \bibinfo{author}{Marzari, N.} \&
  \bibinfo{author}{Vanderbilt, D.}
\newblock \bibinfo{title}{{Maximally localized Wannier functions for entangled
  energy bands}}.
\newblock \emph{\bibinfo{journal}{Phys. Rev. B}} \textbf{\bibinfo{volume}{65}},
  \bibinfo{pages}{035109} (\bibinfo{year}{2001}).

\bibitem{mostofi2008}
\bibinfo{author}{Mostofi, A.~A.} \emph{et~al.}
\newblock \bibinfo{title}{{wannier90: A tool for obtaining maximally-localised
  Wannier functions}}.
\newblock \emph{\bibinfo{journal}{Comp. Phys. Comm.}}
  \textbf{\bibinfo{volume}{178}}, \bibinfo{pages}{685--699}
  (\bibinfo{year}{2008}).

\bibitem{wang2006}
\bibinfo{author}{Wang, X.}, \bibinfo{author}{Yates, J.},
  \bibinfo{author}{Souza, I.} \& \bibinfo{author}{Vanderbilt, D.}
\newblock \bibinfo{title}{{Ab initio calculation of the anomalous Hall
  conductivity by Wannier interpolation}}.
\newblock \emph{\bibinfo{journal}{Phys. Rev. B}} \textbf{\bibinfo{volume}{74}},
  \bibinfo{pages}{195118} (\bibinfo{year}{2006}).

\end{thebibliography}
\end{document}